\documentstyle[11pt]{article}
\begin{document}

\title{\textbf{Torsional Alfv\'{e}n waves in stratified and expanding magnetic flux
tubes}}
\author{K. Karami$^{1}$\thanks{E-mail: KKarami@uok.ac.ir} ,
K. Bahari${^2}\thanks{E-mail:
K$_{-}$Bahari@iasbs.ac.ir}$\\$^{1}$\small{Department of Physics,
University of Kurdistan, Pasdaran Street, Sanandaj,
Iran}\\$^{2}$\small{Institute for Advanced Studies in Basic Sciences
(IASBS), Gava Zang, Zanjan, Iran}}

\maketitle

\begin{abstract}
The effects of both density stratification and magnetic field
expansion on torsional Alfv\'{e}n waves in magnetic flux tubes are
studied. The frequencies, the period ratio $P_1/P_2$ of the
fundamental and its first-overtone, and eigenfunctions of torsional
Alfv\'{e}n modes are obtained. Our numerical results show that the
density stratification and magnetic field expansion have opposite
effects on the oscillating properties of torsional Alfv\'{e}n waves.
\end{abstract}
\noindent{{\bf Keywords:}~~~Sun: corona . Sun: magnetic fields .
Sun: oscillations}
\clearpage
\section{Introduction}\label{intro}
Hannes Alfv\'{e}n (1942) predicted the existence of Alfv\'{e}n wave
which is one of the magnetohydrodynamic (MHD) waves propagating in
magnetized plasmas such as the solar atmosphere. Torsional
Alfv\'{e}n waves can be observed as temporal and spatial variations
in spectral emission along the coronal loops (Zaqarashvili 2003).
They are an ideal tool for coronal seismology as their phase speed
depends on plasma quantities within the loop alone, while wave
speeds of magnetosonic oscillations are influenced by plasma
conditions in the ambient medium (Zaqarashvili \& Murawski 2007).
More recently, torsional Alfv\'{e}n waves in the solar atmosphere
were discovered by Jess et al. (2009) using the high-resolution
Swedish Solar Telescope.

Torsional Alfv\'{e}n modes can have a significant role in coronal
heating and solar wind acceleration, based upon the ability of
torsional waves to penetrate easily into the corona (see e.g.
Ruderman 1999; Copil, Voitenko \& Goossens 2008). Possible
Alfv\'{e}n wave dissipation processes that may be responsible for
heating the solar plasma have been investigated by various authors,
e.g., dissipation in resonant layers (Poedts, Goossens \& Kerner
1989, 1990a,b,c; Erd\'{e}lyi \& Goossens 1995; Ruderman et al.
1997a,b; Erd\'{e}lyi 1998; Ruderman 1999; Narain et al. 2001; Safari
et al. 2006; Karami, Nasiri \& Amiri 2009; Karami \& Bahari 2010)
and phase mixing (Heyvaerts \& Priest 1983; Ruderman, Nakariakov \&
Roberts 1998; Smith, Tsiklauri \& Ruderman 2007; Karami \& Ebrahimi
2009).

People have also paid special attention to the nonlinear effects of
torsional Alfv\'{e}n modes. It was demonstrated numerically that the
observed spiky intensity profiles due to impulsive energy releases
could be obtained from nonlinear torsional waves (see e.g. Moriyasu
et al. 2004; Antolin et al. 2008). Taroyan (2009) showed that
small-amplitude Alfv\'{e}n waves can be amplified into the nonlinear
regime by the presence of siphon flows in coronal loops.

Zaqarashvili \& Murawski (2007) investigated the evolution of
torsional Alfv\'{e}n waves in longitudinally inhomogeneous coronal
loops. They concluded that the inhomogeneous mass density field
leads to the reduction of a wave frequency of torsional
oscillations, in comparison to that estimated from mass density at
the loop apex. Also this frequency reduction results from the
decrease of an average Alfv\'{e}n speed as far as the inhomogeneous
loop is denser at its footpoints.

Copil, Voitenko \& Goossens (2010) studied torsional Alfv\'{e}n
waves in twisted small scale current threads of the solar corona.
They showed that the trapped Alfv\'{e}n eigenmodes do exist and are
localized in thin current threads where the magnetic field is
twisted. They pointed out that the wave spectrum is discrete in
phase velocity, and the number of modes is finite and depends on the
amount of the magnetic field twist. Also the phase speeds of the
modes are between the minimum of the Alfv\'{e}n speed in the
interior and the exterior Alfv\'{e}n speed.

Vasheghani Farahani, Nakariakov \& Van Doorsselaere (2010)
investigated torsional axisymmetric long wavelength MHD modes of
solar coronal plasma structures with the use of the second order
thin flux tube approximation. They concluded that the phase speed of
torsional waves depends upon the direction of the wave propagation,
and also the waves are compressible.

Verth, Erd\'{e}lyi \& Goossens (2010) studied the observable
properties of torsional Alfv\'{e}n waves in both thin and
finite-width stratified and expanding magnetic flux tubes. They
demonstrated that for thin flux tubes, observation of the eigenmodes
of torsional Alfv\'{e}n waves can provide temperature diagnostics of
both the internal and surrounding plasma. They also showed that in
the finite-width flux tube regime, these waves are the ideal
magneto-seismological tool for probing radial plasma inhomogeneity
in solar waveguides.

All mentioned in above motivate us to have further investigates on
torsional Alfv\'{e}n waves by considering the effects of both
density stratification and magnetic field expansion on the
frequencies and eigenfunctions of torsional Alfv\'{e}n modes in the
magnetic flux tubes. This paper is organized as follows. In Section
2 we introduce the model and derive the equations of motion. In
Section 3 we give numerical results. Section 4 is devoted to
conclusions.

\section{Model and equations of motion}
We consider an expanding magnetic flux tube of length $2L$ with
longitudinal plasma density as typical coronal loop. The tube is
assumed to be thin, $r_a/L\ll 1$, where $r_a$ is the tube radius at
the apex. Following Ruderman, Verth $\&$ Erd\'{e}lyi (2008) and
Verth $\&$ Erd\'{e}lyi (2008), the background magnetic field is
assumed to has both radial and axial components with $r$- and
$z$-dependence, i.e. $B_r=B_r(r,z)$ and $B_z=B_z(r,z)$. The coronal
plasma is nearly zero-$\beta$ and this yields the magnetic field to
be force free. For the selected magnetic field, the electrical
current is in the $\phi$-direction. Hence the force free condition,
i.e. ${\mathbf{J}}\times{\mathbf B}=0$, is satisfied when the
electrical current ${\mathbf{J}}=\nabla\times{\mathbf B}=0$. The
background magnetic field can be related to a vector potential field
${\mathbf A}$ as
\begin{equation}
{\bf B}=\nabla\times{\bf A},\label{b}
\end{equation}
where
\begin{equation}
{\mathbf A}=\frac{\psi(r,z)}{r}~{\mathbf e}_\phi.\label{a}
\end{equation}
Therefore the radial and axial components of the magnetic field can
be expressed in terms of the scalar potential $\psi$ as
\begin{equation}
B_r=\frac{-1}{r}\frac{\partial\psi}{\partial
z},~~~B_z=\frac{1}{r}\frac{\partial\psi}{\partial
r},\label{bcomponents}
\end{equation}
where the magnetic field here is perpendicular to $\nabla\psi(r,z)$,
i.e. the magnetic field lines lie in the surface $\psi(r,z)=$
constant. Hence, the equation of the tube boundary is given by
$\psi(r,z)=\psi_0$, where $\psi_0$ is a constant. Here, our aim is
to study the torsional Alfv\'{e}n waves in which the surface of the
flux tube $\psi=\psi_0$ has an oscillating motion in the azimuthal
direction. If we apply the force free condition $\nabla\times{\bf
B}=0$ to the equilibrium magnetic field (\ref{b}), we obtain a
partial differential equation for $\psi$ as
\begin{eqnarray}
\frac{\partial^2\psi}{\partial
r^2}-\frac{1}{r}\frac{\partial\psi}{\partial r}
+\frac{\partial^2\psi}{\partial z^2}=0\label{psi}.
\end{eqnarray}
Verth $\&$ Erd\'{e}lyi (2008) solved the above equation and found
the $z$-component of the magnetic field as
\begin{eqnarray}
B_z(z)=B_{z,f}\left\{1+\frac{(1-\Gamma^2)}{\Gamma^2}\frac{\Big[\cosh\big(\frac{z}{L}\big)-\cosh(1)\Big]}
{1-\cosh(1)}\right\},\label{bz}
\end{eqnarray}
and the radius of the magnetic flux tube boundary as
\begin{eqnarray}
r(z)=r_f\left\{1+\frac{(1-\Gamma^2)}{\Gamma^2}\frac{\Big[\cosh(\frac{z}{L})
-\cosh(1)\Big]}{1-\cosh(1)}\right\}^{-1/2},\label{r0}
\end{eqnarray}
where $B_{z,f}=B_z(\pm L)$ and $r_f=r(\pm L)$ are $z$-component of
the magnetic field and the radius of flux tube at the loop
footpoints, respectively. Also
$\Gamma=\frac{r_a}{r_f}=\frac{r(0)}{r(\pm L)}$ is the tube expansion
factor which is defined as ratio of the tube radius at the apex
($z=0$) to the tube radius at the footpoints ($z=\pm L$). For a tube
with constant cross section, the expansion factor is unity but for
an expanding flux tube we have $\Gamma>1$. As Ruderman, Verth $\&$
Erd\'{e}lyi (2008) emphasized the important property of this
particular model is that it can describe only magnetic tubes with
relatively small expansion factors, definitely smaller than $1.87$.
We also take into account the effect of density stratification and
assume that the density varies exponentially with the height, $h$,
in the atmosphere as $\rho=\rho_fe^{-\frac{h}{H}}$. Here $\rho_f$ is
the density at the footpoints and $H$ is the density scale height.
Following Ruderman, Verth \& Erd\'{e}lyi (2008) for a half-circle
loop with length $2L$ the density can be written as
\begin{eqnarray}
\rho(z)=\rho_f\exp{\left[-2\mu \cos\left(\frac{\pi
z}{2L}\right)\right]},~~~\mu:=\frac{L}{\pi H}, \label{rho}
\end{eqnarray}
where $\mu$ is defined as stratification parameter.

Note that the Alfv\'{e}n velocity in our selected model varies only
along the background magnetic field. Since we used the thin tube
approximation, hence the variation across the magnetic field lines
is negligible. Ruderman et al. (1997a,b) to study the resonant
absorption of torsional Alfv\'{e}n waves considered the variation of
Alfv\'{e}n velocity only across the magnetic field.

The linearized MHD equations for a zero-$\beta$ plasma are
\begin{eqnarray}
\frac{\partial^2\bf{\xi}}{\partial
t^2}=\frac{1}{4\pi\rho}\big(\nabla\times\delta\bf{B}\big)\times\bf{B},\label{mhd1}
\end{eqnarray}
\begin{eqnarray}
\delta\bf{B}=\nabla\times\big(\bf{\xi}\times B\big),\label{mhd2}
\end{eqnarray}
where ${\bf \xi}=(0,0,\xi_{\phi})$ is the Lagrangian displacement of
the plasma and $\delta{\bf{B}}=(0,0,\delta B_{\phi})$ is the
Eulerian perturbation in the magnetic field. Note that in Eq.
(\ref{mhd1}) due to the force free background magnetic field, the
term
$\frac{1}{4\pi\rho}\big(\nabla\times\bf{B}\big)\times\delta\bf{B}$
is absent.

We rewrite Eqs. (\ref{mhd1}) and (\ref{mhd2}) in components as
\begin{eqnarray}
\frac{\partial^2\xi_{\phi}}{\partial
t^2}=\frac{1}{4\pi\rho}\left[\frac{B_r}{r}\frac{\partial(r\delta
B_{\phi})}{\partial r}+B_z\frac{\partial\delta B_{\phi}}{\partial
z}\right],\label{xiphi1}
\end{eqnarray}
\begin{eqnarray}
\delta B_{\phi}=\frac{\partial(B_r\xi_{\phi})}{\partial r}
+\frac{\partial(B_z\xi_{\phi})}{\partial z}.\label{bphi1}
\end{eqnarray}
Now like Ruderman, Verth \& Erd\'{e}lyi (2008) and Verth,
Erd\'{e}lyi \& Goossens (2010) we use a non-orthogonal flux
coordinate system in which $\psi$ becomes an independent variable
instead of $r$, i.e. $r=r(\psi,z)$. In this coordinate system, an
arbitrary function $f(r,z)$ will be transformed to another function
$F(\psi,z)$ as $f(r,z)=F(\psi(r,z),z)$. Therefore, using Eq.
(\ref{bcomponents}) the $r$ and $z$ partial derivatives of $f$
transform to
\begin{eqnarray}
\Big(\frac{\partial f}{\partial r}\Big)_z=rB_z\frac{\partial
F}{\partial\psi},~~~\Big(\frac{\partial f}{\partial
z}\Big)_\psi=\frac{\partial F}{\partial z}-rB_r\frac{\partial
F}{\partial\psi}.\label{dif1}
\end{eqnarray}
Differentiating the identities $\psi=\psi\Big(r(\psi,z),z\Big)$ and
$r=r\Big(\psi(r,z),z\Big)$ with respect to $z$ and using Eq.
(\ref{bcomponents}) one can get
\begin{eqnarray}
\frac{\partial r}{\partial z}=\frac{B_r}{B_z},~~~\frac{\partial
r}{\partial\psi}=\frac{1}{rB_z}.\label{dif2}
\end{eqnarray}
Using Eqs. (\ref{dif1}) and (\ref{dif2}) one can rewrite the field
Eqs. (\ref{xiphi1}) and (\ref{bphi1}) as
\begin{eqnarray}
\frac{\partial^2\xi_{\phi}}{\partial t^2}=\frac{B_z}{4\pi
r\rho}\frac{\partial(r\delta B_{\phi})}{\partial z}, \label{xiphi2}
\end{eqnarray}
\begin{eqnarray}
\delta B_{\phi}=rB_z\frac{\partial}{\partial
z}\Big(\frac{\xi_{\phi}}{r}\Big).\label{bphi2}
\end{eqnarray}
Substituting Eq. (\ref{bphi2}) into (\ref{xiphi2}) and considering
the time-dependence as $e^{-i\omega t}$ the result yields
\begin{eqnarray}
\frac{B_z}{4\pi r}\frac{\partial}{\partial
z}\left[r^2B_z\frac{\partial}{\partial
z}\left(\frac{\xi_{\phi}}{r}\right)\right]+\rho\omega^2\xi_{\phi}=0,\label{basic}
\end{eqnarray}
where the variables $B_z(z)$, $r(z)$ and $\rho(z)$ are given by Eqs.
(\ref{bz}), (\ref{r0}) and (\ref{rho}), respectively. Equation
(\ref{basic}) is the same as Eq. (24) in Verth, Erd\'{e}lyi \&
Goossens (2010) and also the same as the equation for continuum
Alfv\'{e}n waves on individual magnetic surfaces for a purely
poloidal magnetic field for any azimuthal wave number $m$ (see Eq.
71 in Poedts, Hermans \& Goossens 1985).

Following Verth, Erd\'{e}lyi \& Goossens (2010) in the thin flux
tube limit, all magnetic surfaces oscillate with the same frequency,
i.e. the frequency is independent of magnetic surface of constant
$\psi$ as long as $r_a/L\ll 1$. Therefore, the torsional Alfv\'{e}n
waves described by Eq. (\ref{basic}) are global in which the whole
loop oscillates with one frequency. For the finite width flux tube,
Verth, Erd\'{e}lyi \& Goossens (2010) found the local torsional
Alfv\'{e}n wave frequencies that vary across the magnetic surfaces.

One notes that the MHD waves have mixed properties. For instance,
torsional Alfv\'{e}n waves can be coupled to kink waves (see e.g. De
Groof \& Goossens 2000, 2002; Goossens \& De Groof 2001; De Groof,
Paes \& Goossens 2002;  Goossens, De Groof \& Andries 2002;
Goossens, Andries \& Arregui 2006; Goossens 2008). Indeed, the pure
torsional Alfv\'{e}n waves require an axisymmetric background with a
purely poloidal magnetic field (no azimuthal component) and
axisymmetric motions.

In the next section we solve the above differential equation using
the suitable boundary conditions to obtain the eigenvalues $\omega$
and the eigenfunctions $\xi_{\phi}(z)$.

\section{Numerical results}
Here, we solve Eq. (\ref{basic}) using the shooting method to obtain
both the eigenfrequencies and eigenfunctions of torsional Alfv\'{e}n
waves in stratified and expanding magnetic flux tube. We use the
rigid boundary conditions and assume that
$\xi_{\phi}(-L)=\xi_{\phi}(L)=0$. As typical parameters for a
coronal loop, we assume $2L = 10^5~{\rm km}$,
$\rho_f=2\times10^{-14}~{\rm gr~cm^{-3}}$ and $B_{z,f} = 100~{\rm
G}$. For such a loop, one finds Alfv\'{e}n speed
$v_A=\frac{B_{z,f}}{\sqrt{4\pi\rho_f}}=2000~{\rm km~s^{-1}}$ and
Alfv\'{e}n frequency $\omega_A:=\frac{v_A}{2L}=0.02~{\rm
rad~s^{-1}}$.

The effects of magnetic field expansion and density stratification
on the frequencies of the fundamental and first-overtone $n=1,2$
torsional Alfv\'{e}n modes are displayed in Figs. \ref{wgama} and
\ref{wmu}, respectively. Figure \ref{wgama} shows that for a given
stratification parameter $\mu$, the frequencies of the fundamental
and first-overtone modes decrease with increasing the expansion
factor $\Gamma$. This result is in agreement with that obtained by
Ruderman, Verth \& Erd\'{e}lyi (2008) for the kink modes. Figure
\ref{wmu} presents that for a given expansion factor $\Gamma$, the
frequencies of the fundamental and first-overtone modes increase
when the stratification parameter increases. This also is in good
concord with the result derived by Karami, Nasiri \& Amiri (2009)
for the kink and fluting body waves. One notes that Verth,
Erd\'{e}lyi \& Goossens (2010) showed that for a vertical stratified
and expanding thin magnetic flux tube, for the isothermal case in
which the Alfv\'{e}n velocity is constant along the magnetic field,
the frequency of torsional Alfv\'{e}n waves remains unchanged. But
for the non-isothermal case in which the Alfv\'{e}n velocity varies
along the flux tube, the frequency of oscillations for hot and cool
tubes decreases and increases, respectively. The results of Verth,
Erd\'{e}lyi \& Goossens (2010) for hot and cool tubes are in good
agreement with our results presented in Figs. \ref{wgama} and
\ref{wmu}, respectively. In the other words, one can say that the
cool and hot loops behave like those tubes in which the effect of
density stratification and magnetic field expansion is dominant,
respectively.

The period ratio $P_1/P_2$ of the fundamental and first-overtone
$n=1,2$ torsional Alfv\'{e}n modes versus the expansion factor and
stratification parameter is plotted in Figs. \ref{pgama} and
\ref{pmu}, respectively. Note that the period ratio is used as a
seismological tool to investigate e.g., longitudinal structure
(Andries et al. 2005; Andries, Arregui \& Goossens 2005) and radial
structure (Verth, Erd\'{e}lyi \& Goossens 2010) of magnetic loops.
Figure \ref{pgama} reveals that for a given $\mu$, the period ratio
$P_1/P_2$ increases with increasing the expansion factor. This is in
agreement with the result obtained by Verth \& Erd\'{e}lyi (2008)
for the kink body modes. Figure \ref{pmu} illustrates that for a
given $\Gamma$, the period ratio $P_1/P_2$ decreases when the
stratification parameter increases. This also is in good concord
with the result reported by Karami, Nasiri \& Amiri (2009) for the
kink and fluting body waves. Note that Verth, Erd\'{e}lyi \&
Goossens (2010) pointed out that for a vertical stratified and
expanding thin magnetic flux tube, the period ratio of torsional
Alfv\'{e}n modes for the isothermal case is 2. Whereas for the
non-isothermal case, the period ratio for both cool and hot tubes is
bigger than 2. They also considered a semicircular thin magnetic
flux tube and showed that the period ratio for the isothermal case
is 2. But for the non-isothermal case, the period ratio for hot and
cool tubes becomes bigger and smaller than 2, respectively. The
results of Verth, Erd\'{e}lyi \& Goossens (2010) for hot and cool
semicircular thin tubes are in good agreement with our results
illustrated in Figs. \ref{pgama} and \ref{pmu}, respectively.

From Figs. \ref{wgama} to \ref{pmu} one can conclude that the
magnetic field expansion and density stratification have opposite
effects on the frequencies and period ratio $P_1/P_2$ of torsional
Alfv\'{e}n modes. This is expected because as a result of magnetic
field expansion, Alfv\'{e}n speed becomes smaller at the loop apex
and the travel time of wave between the two footpoints becomes
greater. But density stratification causes Alfv\'{e}n speed becomes
greater at the loop apex. Hence, the travel time of wave between the
two footpoints becomes smaller.

Here, we also investigate the effects of both the magnetic field
expansion and density stratification on the eigenfunctions of
torsional Alfv\'{e}n modes. Figures \ref{antigama} and \ref{antimu}
show the normalized anti-node shift $\Delta z/L$ of the
first-overtone eigenfunction versus the magnetic expansion factor
and stratification parameter, respectively. The anti-node shift of
the first-overtone mode is one of the important quantities in
coronal seismology. For instance, it was pointed out by Verth (2007)
that if the magnetic field becomes weaker towards the loop apex, a
useful magneto-seismological signature is the anti-node of the 1st
harmonic since it shifts towards the loop apex. Also Andries,
Arregui \& Goossens (2009) used the anti-node shift in the presence
of longitudinal density stratification as signature of the eigenmode
modifications. They showed that the anti-node shift of the
eigenfunctions of the higher kink-oscillation overtones are even
more than the first overtone. Figure \ref{antigama} clears that for
a given stratification parameter, the normalized anti-node shift
increases with increasing the expansion factor. This is in good
agreement with the result obtained by Verth \& Erd\'{e}lyi (2008)
for the kink body modes. Figure \ref{antimu} presents that for a
given expansion factor, the normalized anti-node shift decreases
when the stratification parameter increases. In the other words, the
magnetic field expansion causes positive anti-node shift and density
stratification causes negative anti-node shift. To illustrate this
in more detail, the eigenfunctions of the first-overtone torsional
Alfv\'{e}n modes for the different values of $\mu$ and $\Gamma$ are
displayed in Fig. \ref{eigenfunction}. It shows that as a result of
density stratification, the anti-nodes of the first-overtone
eigenfunction shift away from the loop apex ($z=0$). This result was
also obtained by Andries, Arregui \& Goossens (2009) for kink modes.
Figure \ref{eigenfunction} clears also that in the presence of
magnetic field expansion, the anti-nodes shift towards the loop apex
which is in agreement with the result obtained by Verth \&
Erd\'{e}lyi (2008) for the kink body modes. It is remarkable that
Verth, Erd\'{e}lyi \& Goossens (2010) illustrated that for a
vertical stratified and expanding thin magnetic flux tube, the
antinodes of the eigenfunctions for the isothermal case remains
unshifted. But for the non-isothermal case for a cool tube, the
distance between the antinodes becomes more spaced out with height
and for a hot tube the antinodes become closer together with height.
The results of Verth, Erd\'{e}lyi \& Goossens (2010) for cool and
hot tubes are in agreement with our results displayed in the up and
down panels of Fig. \ref{eigenfunction}, respectively.

Note that in some cases in above we compared our results for the
frequencies, period ratio and eigenfunctions of torsional Alfv\'{e}n
modes with those obtained by others for kink modes. Although
torsional Alfv\'{e}n waves ($m$ = 0) and kink waves ($m$ = 1) have
axisymmetric and nonaxisymmetric motions, respectively, in the
presence of the magnetic field expansion and density stratification
they show the similar behaviour. This may be caused by this fact
that the restoring force in both of them is the magnetic tension
force. This is why that Goossens et al. (2009) used the adjective
Alfv\'{e}nic for kink modes.

\section{Conclusions}\label{Con}
Here, the effects of density stratification and magnetic filed
expansion on torsional Alfv\'{e}n waves in coronal loops are
studied. To do this, a typical coronal loop is considered as an
expanding magnetic flux tube that undergoes a density varying along
the tube. The linearized MHD equations are reduced to an eigenvalue
problem for the azimuthal component of the Lagrangian displacement
of the plasma. Using the shooting method and under the rigid
boundary conditions for the loop footpoints, both the
eigenfrequencies and eigenfunctions of the fundamental and
first-overtone torsional Alfv\'{e}n modes are obtained. Our numerical results show the following.\\

i) For a given density stratification parameter $\mu$, the
frequencies of the fundamental and first-overtone torsional
Alfv\'{e}n modes decrease and the period ratio $P_1/P_2$ increases
when the magnetic field expansion factor $\Gamma$ increases.\\

ii) For a given $\Gamma$, the frequencies of the fundamental and
first-overtone torsional Alfv\'{e}n modes increase and the period
ratio $P_1/P_2$ decreases when $\mu$ increases.\\

iii) Both the density stratification and magnetic field expansion
shift the location of the anti-nodes of the first-overtone torsional
Alfv\'{e}n modes but in the apposite directions.\\

All mentioned in above illustrate that the density stratification
and magnetic field expansion have opposite effects on the
oscillating properties of torsional Alfv\'{e}n waves.
\\
\\
\noindent{\textbf{Acknowledgements}\\
The authors thank the anonymous referee for very valuable comments.

\clearpage
 \begin{figure}
\center \includegraphics{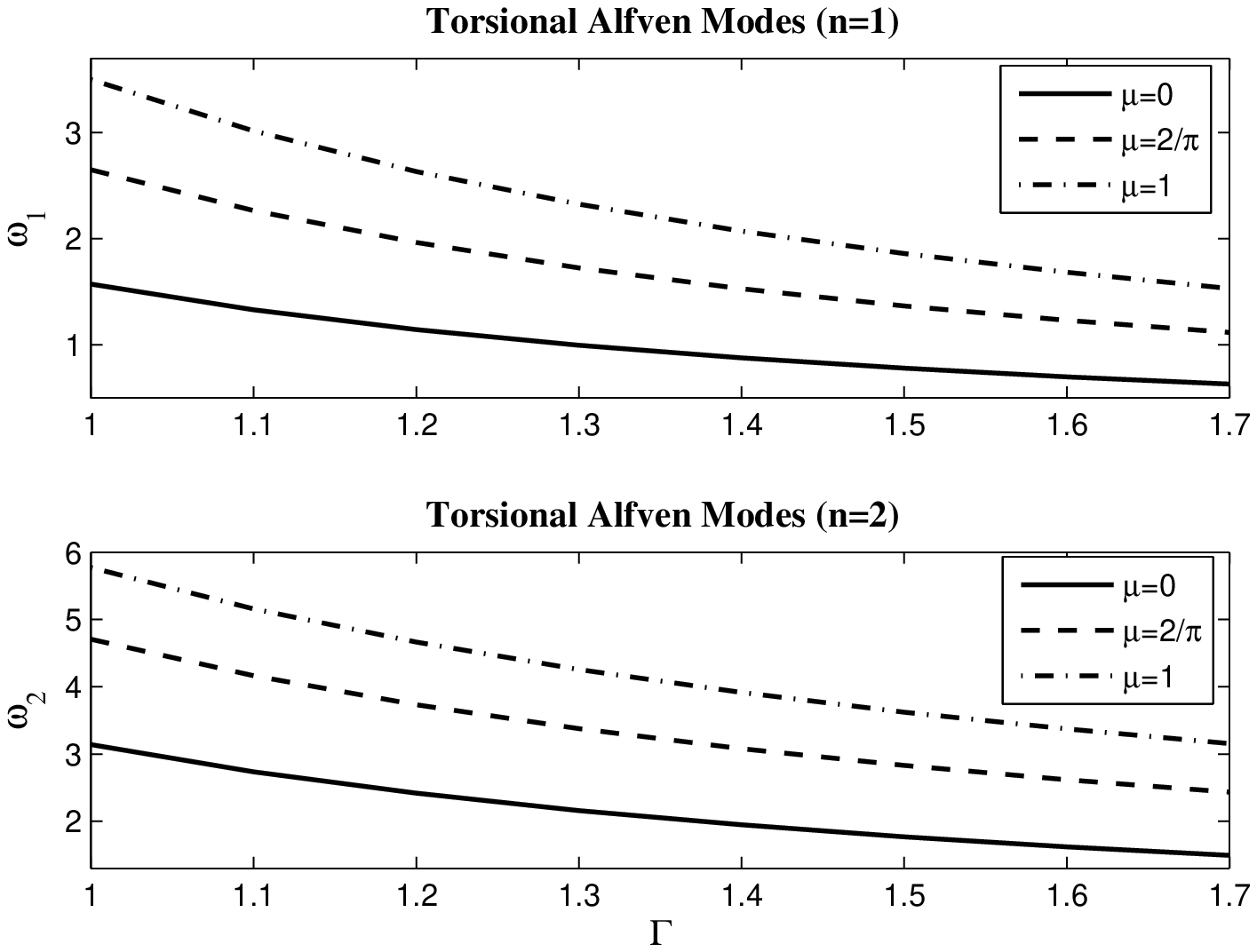}
      \vspace{7.5cm}
      \caption[]{Frequencies of the fundamental and first-overtone torsional Alfv\'{e}n modes versus
      the expansion factor, $\Gamma=\frac{r_a}{r_f}$, for different stratification parameter
      $\mu=$ 0 (solid line), $2/\pi$ (dashed line) and
1 (dash-dotted line). The loop parameters are: $2L=10^5$ km,
 $\rho_f=2\times 10^{-14}$ gr cm$^{-3}$,
$B_{z,f}=100$ G. Frequencies are in units of Alfv\'{e}n frequency,
$\omega_{\rm A}= 0.02{\rm~rad~s^{-1}}$.}
         \label{wgama}
   \end{figure}
 \begin{figure}
\center \includegraphics{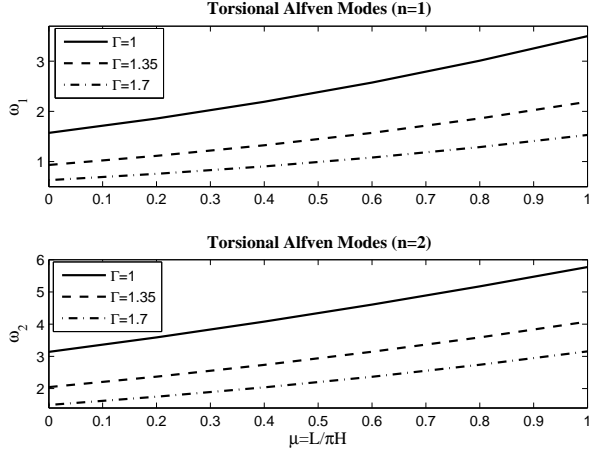}
      \vspace{7.5cm}
      \caption[]{Frequencies of the fundamental and first-overtone torsional Alfv\'{e}n modes
      versus the stratification parameter, $\mu=\frac{L}{\pi H}$, for different expansion factor
      $\Gamma=$ 1 (solid line), 1.35 (dashed line) and 1.7
 (dash-dotted line). Auxiliary parameters as in Fig.
\ref{wgama}.}
         \label{wmu}
   \end{figure}
 \begin{figure}
\center \includegraphics{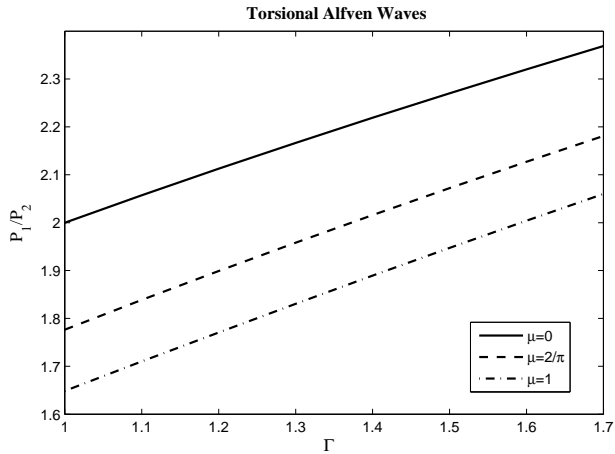}
      \vspace{7.5cm}
\caption[]{The period ratio $P_1/P_2$ of the fundamental and
first-overtone torsional Alfv\'{e}n waves versus the expansion
factor for different stratification parameter $\mu=$ 0 (solid line),
$2/\pi$ (dashed line) and 1 (dash-dotted line). Auxiliary parameters
as in Fig. \ref{wgama}.}
         \label{pgama}
   \end{figure}
 \begin{figure}
\center \includegraphics{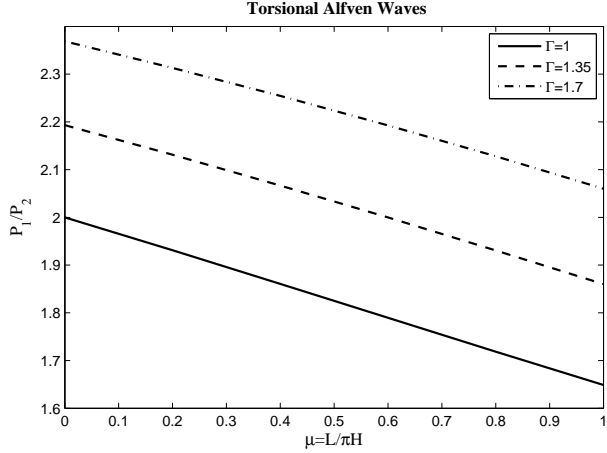}
      \vspace{7.5cm}
\caption[] {The period ratio $P_1/P_2$ of the fundamental and
first-overtone torsional Alfv\'{e}n waves versus the stratification
parameter for different expansion factor $\Gamma=$ 1 (solid line),
1.35 (dashed line) and 1.7 (dash-dotted line). Auxiliary parameters
as in Fig. \ref{wgama}.}
         \label{pmu}
   \end{figure}
 \begin{figure}
\center \includegraphics{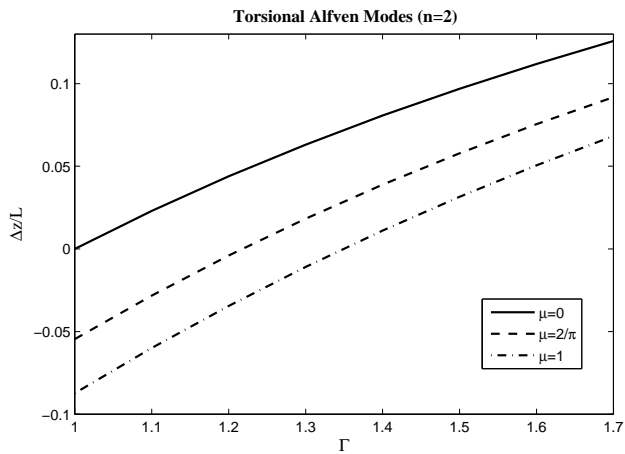}
      \vspace{7.5cm}
\caption[] {Normalized shift of the first-overtone anti-node versus
the expansion factor for different stratification parameter $\mu=$ 0
(solid line), $2/\pi$ (dashed line) and 1 (dash-dotted line).
Auxiliary parameters as in Fig. \ref{wgama}.}
         \label{antigama}
   \end{figure}
 \begin{figure}
\center \includegraphics{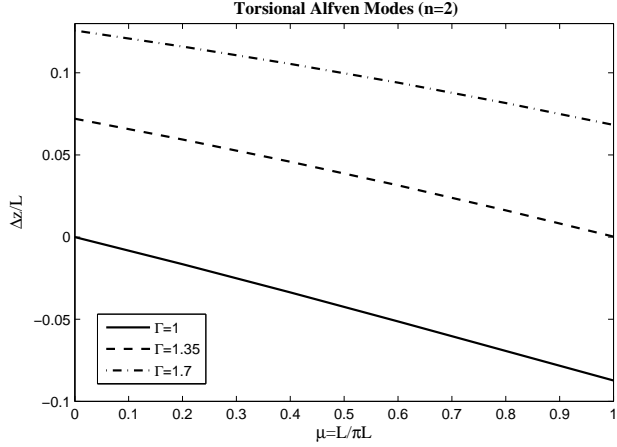}
      \vspace{7.5cm}
\caption[] {Normalized shift of the first-overtone anti-node versus
the stratification parameter for different expansion factor
$\Gamma=$ 1 (solid line), 1.35 (dashed line) and 1.7 (dash-dotted
line). Auxiliary parameters as in Fig. \ref{wgama}.}
         \label{antimu}
   \end{figure}
 \begin{figure}
\center \includegraphics{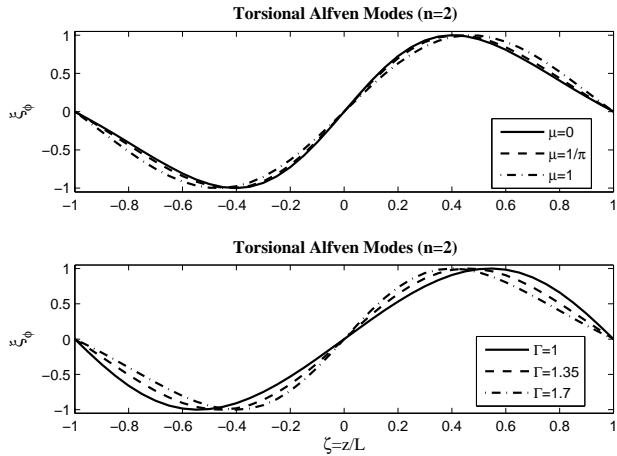}
      \vspace{7.5cm}
\caption[] {Eigenfunctions of the first-overtone torsional
Alfv\'{e}n modes against fractional length $\zeta=z/L$ for different
expansion factor and stratification parameter. In the up panel
$\Gamma=1.5$ and in the down panel $\mu=0.5$. Auxiliary parameters
as in Fig. \ref{wgama}.}
         \label{eigenfunction}
   \end{figure}
\end{document}